\documentclass[letterpaper,10pt]{article} 
\usepackage{opticameet3} %% use version 3 for proper copyright statement
\usepackage{wrapfig}
\usepackage{subcaption}

\newcommand\authormark[1]{\textsuperscript{#1}}
\providecommand{\passage}{\textsc{Lightpath}\xspace}

\providecommand{\allreduce}{\textsc{AllReduce}\xspace}
\providecommand{\allgather}{\textsc{AllGather}\xspace}
\providecommand{\dandc}{\textsc{D\&C}\xspace}

\usepackage{amsmath,amssymb}
\usepackage[colorlinks=true,bookmarks=false,citecolor=blue,urlcolor=blue]{hyperref} %pdflatex
\usepackage{subcaption}
\usepackage{graphicx}
\usepackage{xspace}
\usepackage{wrapfig}

\providecommand{\ie}{\emph{i.e.,} }
\providecommand{\eg}{\emph{e.g.,} }

\providecommand{\myparab}[1]{\vspace{1pt}\noindent\textbf{#1} }

\providecommand{\sysname}{\textsc{Lumorph}\xspace}
\providecommand{\passage}{\textsc{Lightpath}\xspace}

\providecommand{\allreduce}{\textsc{AllReduce}\xspace}
\providecommand{\allgather}{\textsc{AllGather}\xspace}

\providecommand{\reducescatter}{\textsc{ReduceScatter}\xspace}
\providecommand{\dandc}{\textsc{D\&C}\xspace}

\begin{document}

\title{Chip-to-chip photonic connectivity in multi-accelerator servers for ML}

\author{Abhishek Vijaya Kumar\authormark{1,*}, Arjun Devraj\authormark{1},  Darius Bunandar\authormark{2}, and Rachee Singh\authormark{1}}

\address{\authormark{1} Cornell University, \authormark{2} Lightmatter}
\vspace{-1mm}
\email{\authormark{*}abhishek@cs.cornell.edu} %% email address is required

\vspace{-2em}

\begin{abstract*}
We present a rack-scale compute architecture for ML using multi-accelerator servers connected via chip-to-chip silicon photonic components. Our architecture achieves (1) multi-tenanted resource slicing without fragmentation, (2) 74\% faster rack-scale collective communication, and (3) $1.7$X speedup in end-to-end ML training throughput.
\end{abstract*}
\vspace{-1pt}

\section{Introduction}

As the demand for computational power for machine learning (ML) rapidly grows, interconnecting accelerators without starving them for data is crucial. 
Traditional electrical interconnects are insufficient due to their large size and high power consumption. Photonic interconnects, leveraging wavelength division multiplexing (WDM), offer a promising solution to meet these density requirements of inter-accelerator communication within multi-accelerator servers. While most research efforts focus on improving transceiver bandwidth, we address the challenge from an interconnect architecture perspective. Specifically, we propose a novel algorithm that integrates optical transceivers and circuit switching to accelerate ML training.

Collective communication primitives, such as \allreduce, are on the critical path of ML training and inference. Recent work has developed optically reconfigurable datacenter fabrics~\cite{karen_bcube, TPUv4} that can allocate direct-connect topologies \ie topologies with point-to-point optical connections between accelerators to achieve contention-free collective communication. However, such fabrics can be susceptible to \emph{compute fragmentation}: a phenomenon where an existing allocation of compute resources to tenants makes it impossible to allocate direct-connect topologies of remaining resources to new tenants. Simply reconfiguring the network fabric (\eg optical switches, transceivers) may not address the resource fragmentation problem.

\myparab{\sysname: an optically reconfigurable rack.}
We address resource fragmentation in multi-tenant ML clusters by developing a new architecture of an optically reconfigurable datacenter rack, called \sysname. Our design connects GPUs within multi-GPU servers with an optically-switched chip-to-chip photonic fabric, \passage.\passage-enabled multi-GPU servers connect to each other via direct attached fibers. \sysname brings optical switching capability close to individual GPUs in servers, enabling any size accelerator allocation connected via direct-connect topologies. The system is a promising solution for AI compute and we demonstrate it using experiments on a lab prototype of \passage. We adapt the recursive doubling/halving algorithm for optimal \allreduce performance on \sysname and show resulting ML throughput gains.

\section{Proposed circuit-switched rack architecture}

% \begin{wrapfigure}{r}{0.4\textwidth}
%   \begin{center}
%       \includegraphics[width=0.4\textwidth]{figures/ofc-fabric.pdf}
%   \end{center}
%   \vspace{-1em}
%   \caption{Server-scale photonic fabric.}
%   \label{fig:fabric} 
%   \vspace{-1em}
% \end{wrapfigure}

\begin{wrapfigure}{r}{0.63\textwidth}
  \centering
  \vspace{-1em}
    \includegraphics[width=0.63\textwidth]{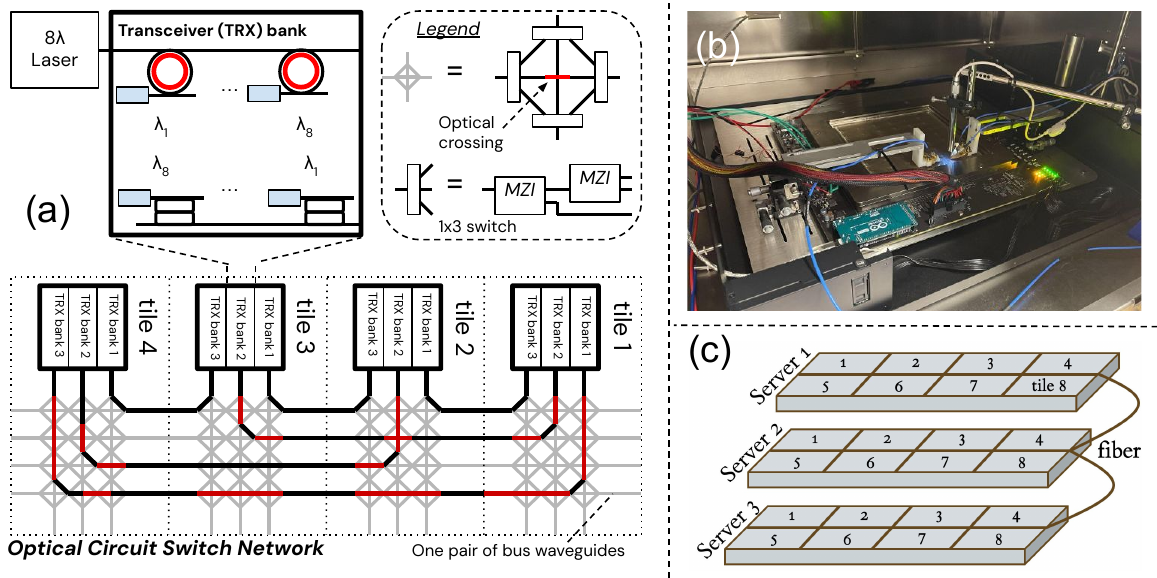}
    \vspace{-2em}
  \caption{Server-scale photonic fabric.} 
  \label{fig:mega1}
\vspace{-2em}
\end{wrapfigure}

\passage is a switchable server-scale photonic fabric in a hybrid Complementary Metal 
Oxide Semiconductor (CMOS) photonics process. A \passage wafer has $32$ or fewer \emph{tiles}.
These tiles are placeholders to 3D stack compute chips (\eg GPUs, CPUs) on the interconnect. A \passage tile has multiple Transmitter and Receiver (TRX) banks. The transmitter uses micro-ring resonators (MRRs) to modulate light. The receiver demultiplexes wavelengths of light, 
converts modulated wavelengths back to electronic data using photodetectors and sends them to the Serializer/Deserializer (SerDes) module. A \passage tile has up to $16$ wavelength-multiplexed lasers.  Waveguides transport wavelengths across tiles.

Each \passage tile has MZI-based optical switches of $1X3$ degree. The hardware enables 
programming the switching behavior of MZIs to implement circuits between accelerators 
on the server. Figure~\ref{fig:mega1}(a) shows one such configuration of the circuit-switched MZI network on \passage that achieves all to all connectivity between four accelerators, each on its own tile. 

We fabricated a testbed that demonstrates the performance of optical devices on \passage in GlobalFoundries. Figure~\ref{fig:mega1}(b) shows the \passage prototype. Fibers can be attached to each tile to connect waveguides to fibers. Using the attached fibers, \sysname will cascade \passage-networked multi-GPU servers into a rack-scale deployment (Figure~\ref{fig:mega1}(c)). We can use area coupling techniques to further extend the topology.

We performed a transmission loopback experiment where a transmitter in \passage is driven by a Xilinx VCU128 FPGA Evaluation board (with Multilane ML4041-K QSFP28 SMA adapter board) and the received data is sent back to the same FPGA for bit error rate (BER) check. PRBS-7 signals are sent from the FPGA’s 28Gbps SerDes to the optical modulator using a radio-frequency probe. The light signals traverse over four \passage tiles with four separate circuit switched networks, and are detected by a Germanium photodetector, whose signals are looped back to the FPGA Evaluation board. At data rates of 10 Gbps, 15 Gbps, and 20 Gbps, we observed very low bit error rates (after open eyes were confirmed) at $6.96 X 10^{-13}$, $6.62 X 10^{-13}$, and $5.60 X 10^{-14}$, respectively. 
% The limited bandwidth of the output Rx amplifier and balun prevented us from completing the BER check at higher baud rate. The results point to the feasibility of the optical network switching and creating high-speed optical data links with low error rates across a silicon photonic wafer.

% \begin{wrapfigure}{r}{0.5\textwidth}
%   \begin{center}
%       \includegraphics[width=0.5\textwidth]{figures/ofc-rack-testbed.pdf}
%   \end{center}
%   \vspace{-1em}
%   \caption{(a) Our \passage hardware testbed. (b) \sysname: an optically reconfigurable rack composed of \passage-networked multi-GPU servers.}
%   \label{fig:rack} 
%     \vspace{-2em}
% \end{wrapfigure}

 In this testbed, reconfiguring the MZI-based optical switches on \passage takes 3.7 microseconds. We configure MZIs such that a pair of bus waveguides make a direct connection from a transceiver of tile A to a transceiver of tile B. Low-loss optical crossings further enable routing along the tiles. The fast reconfiguration of optical switches allows us to establish on-demand optical circuits between a pair of chips stacked on to \passage. 

\section{Limiting compute fragmentation with \sysname}
\begin{wrapfigure}{r}{0.5\textwidth}
    \vspace{-1em}
  \begin{center}
      \includegraphics[width=0.5\textwidth]{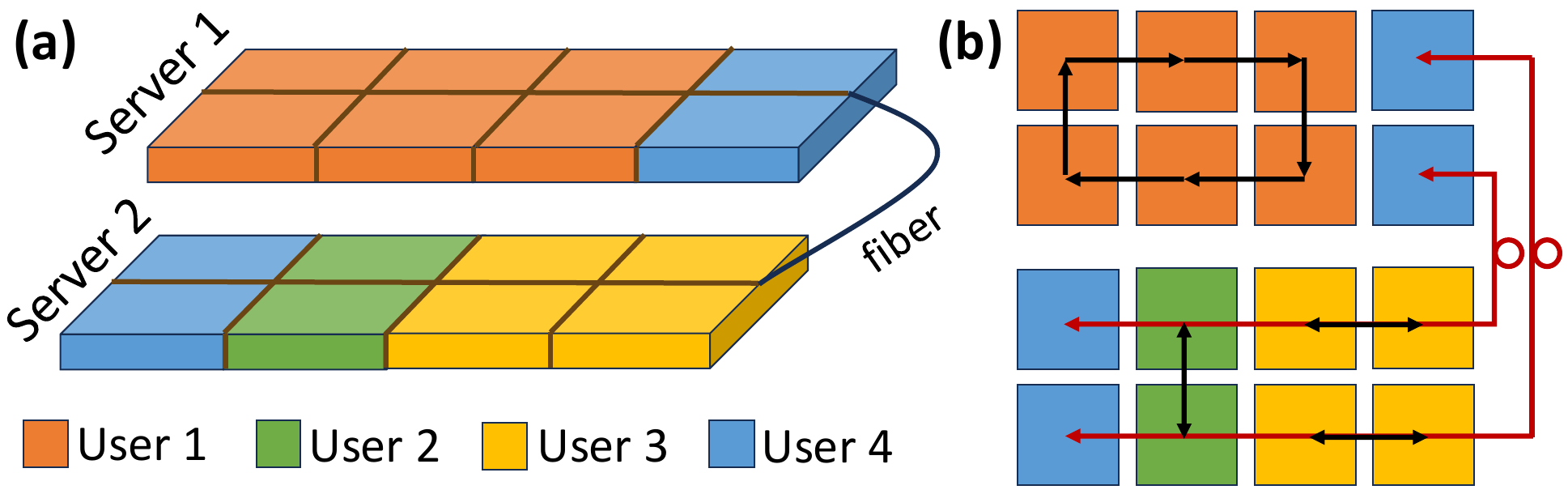}
  \end{center}
  \vspace{-1em}
  \caption{(a) Example allocations in a multi-tenanted environment with fragmented compute. (b) Snapshot of circuits set up for \allreduce in each tenant's allocation.}
  \label{fig:multitenant} 
  \vspace{-1em}
\end{wrapfigure}
\begin{wrapfigure}{r}{0.22\textwidth}
\vspace{-2em}
  \begin{center}
      \includegraphics[width=0.2\textwidth]{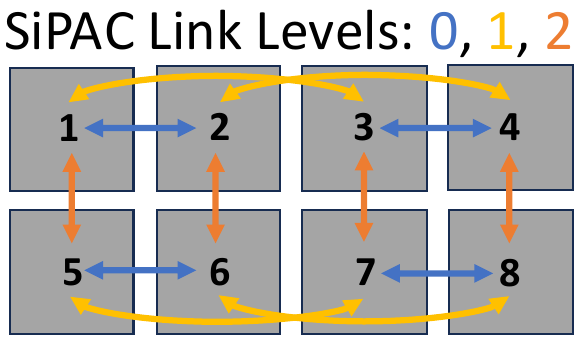}
  \end{center}
  \vspace{-1em}
  \caption{Example configuration with 8 GPUs using \sysname that 
  is equivalent to SiPAC(2,3)~\cite{karen_bcube}.}
  \label{fig:sipac} 
\vspace{-2em}
\end{wrapfigure}

% describe the benefit of reconfiguration for multitenancy and reference the figure
% benefits are not only using the chips, but also doing optimal collective communication
\sysname uses fast reconfiguration of MZIs on \passage and fibers between servers to prevent the fragmentation of compute in multi-tenanted racks. Due to their small pitch, it is feasible to etch thousands of waveguides between tiles on \passage. Using waveguides and MZI configuration, \sysname achieves congestion-free access between any pair of chips in the server. To reach free chips on other servers in the same rack, the circuit can further traverse available fibers. Unlike existing circuit-switched architectures that can only offer compute slice allocations to tenants in fixed and limited sizes (\eg 3D Torus-based TPU~\cite{TPUv4}, BCube-based SiPAC~\cite{karen_bcube}), \sysname can efficiently use datacenter racks for multiple tenants without sacrificing on optimal communication for each tenant. Figure \ref{fig:multitenant}a shows how allocating the request of four chips for User 4 is feasible with \sysname; in existing architectures, this free compute could not be utilized. \sysname can provide arbitrary sized circuit-switched allocations to tenants given enough fibers between servers. \sysname can still use the optimal Flex-SiPCO \allreduce algorithm \cite{karen_bcube}, as tenant topologies can be configured to match the SiPAC topology for any $r$ and $l$ (Figure \ref{fig:sipac}).

Not only does \sysname enable efficient use of rack infrastructure, but it also provides every tenant with the flexibility to configure a topology for optimal \allreduce. Previous work has used \emph{recursive halving} to perform \reducescatter, followed by \emph{recursive doubling} to perform \allgather for optimal \allreduce on a single dimension. While recursive doubling/halving is the optimal \allreduce algorithm when the number of chips is a power of $2$, the popular Ring algorithm achieves bandwidth-optimal \allreduce for other allocation sizes. Figure \ref{fig:multitenant}b shows how User 1 can use Ring \allreduce for their six-chip allocation, while other users can use recursive doubling/halving for their allocations that are a power of $2$. This leverages the reconfigurability of \passage switches both \emph{within the collective} to use recursive doubling/halving \allreduce without wasting NICs and \emph{at the beginning of the job} to configure the optimal \allreduce topology for the specified allocation that uses the ring algorithm.

% We compare \sysname{'s} flexibility in multi-tenanted settings with SiPAC \cite{karen_bcube}, a silicon-photonic circuit-switched architecture for optimal \allreduce. Although SiPAC enables optimal \allreduce for a single workload, it can only provide allocations of fixed, limited sizes that preserve communication optimality for each tenant. Thus, only allocation sizes of $r^{l}$ are feasible for $l \geq 0$ and where $r$ is the optical switch radix in SiPAC. Overall, \sysname{} can theoretically support $100\%$ of possible requested allocation sizes within a rack-scale deployment while SiPAC can only support $2.68\%$. Further, 

\section{Optimal collective communication on \sysname }
The $\alpha$-$\beta$ cost model captures the time taken by collective communication~\cite{taccl}. $\alpha$ represents the fixed cost of sending one data chunk. $beta$ cost represents the data transmission delay. We formalized a general optimization problem to compute the schedule of data transfers that minimize the $\alpha$-$\beta$ cost of collective communication on \sysname, incorporating the cost of MZI reconfiguration.  However, it is challenging to find the global minimum of $\alpha-\beta$ cost because it is a non-convex function of 
the number of links of a GPU. 
% Specifically, if a GPU accesses $L$ lasers with $bw$ bandwidth per laser, the $\alpha$ cost reduces by a factor of $S \leq L$, where $S$ is the number of simultaneous links to other GPUs. However, each link has a bandwidth of 
% $\frac{L*bw}{S}$ so, the $\beta$ cost increases by a factor of $S$. The total cost will be proportional to $(\frac{\alpha}{S} + \beta * S)$. A simple second derivative test shows that the function is non-convex in $S$ and solvers will fail to converge. 
Thus, instead of deriving custom circuit-switching schedules for \sysname, which is computationally intractable, we adapt established algorithms that are known to achieve lower bounds on $\alpha-\beta$ costs, such as recursive doubling/halving. We adapt these algorithms for \sysname, where, instead of always using pre-existing links, \passage establishes on-demand circuits to transfer data. Since establishing circuits incurs a reconfiguration delay, the $\alpha$ cost while using this algorithm on \passage includes the MZI reconfiguration delay. 

Photonic connectivity with \passage allows a GPU to communicate with multiple GPUs in a single round of communication, unlike traditional collective algorithms for fixed network fabrics. \sysname utilizes this to generalize recursive doubling/halving into quadrupling/quartering by splitting a GPU's total egress bandwidth across multiple wavelength-switched circuits. We explore an important tradeoff: splitting a GPU's bandwidth across multiple connections lowers the latency cost ($\alpha$) but raises the bandwidth cost ($\beta$). We explore this tradeoff by evaluating \sysname's doubling/halving (called \sysname-2) and quadrupling/quartering algorithms (\sysname-4).

\begin{wrapfigure}{r}{0.4\textwidth}
\vspace{-2em}
  \begin{center}
      \includegraphics[width=0.4\textwidth]{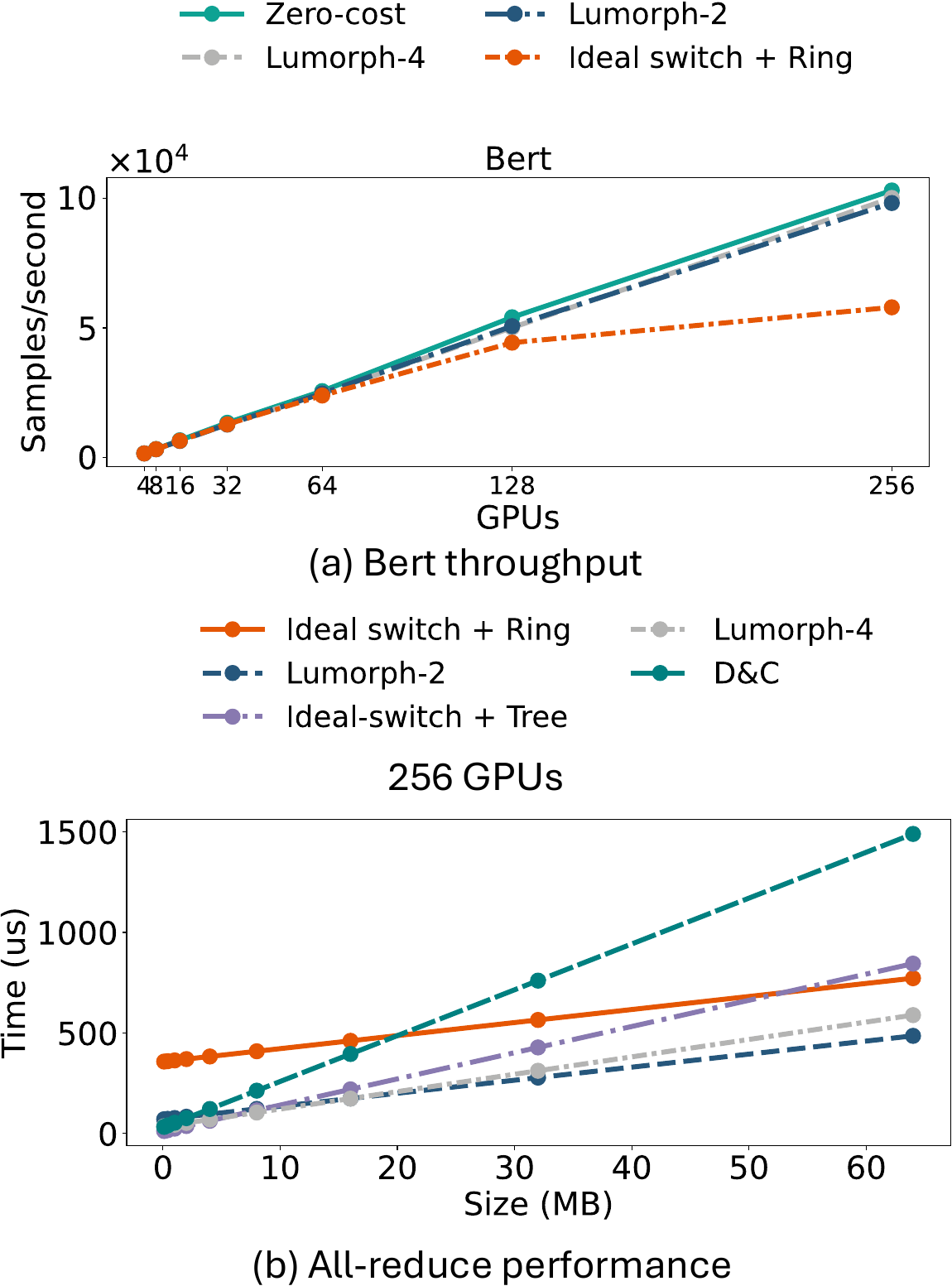}
  \end{center}
  \vspace{-1em}
  \caption{Performance of \sysname.}
  \label{fig:perf} 
\vspace{-2em}
\end{wrapfigure}

We evaluate the performance of \sysname with the adapted \allreduce algorithm (\sysname-2 and \sysname-4) in simulation. We compare \sysname with the hardest baseline --- an electrical interconnect that has an ideal switch with no queuing delays. We note that scaling an electrical packet switched interconnect to 256 GPUs in a rack with no queuing delays is impractical~\cite{taccl}. In addition, we use state-of-the-art Ring and Tree algorithms for \allreduce on the ideal switch-based interconnect implemented by NCCL. \dandc is a greedy solution of the intractable custom collective optimization using a divide and conquer approach. Figure~\ref{fig:perf}(b) shows the run time of collective communication schedules in microseconds as a function of the size of GPU buffers for $64$, $128$ and $256$ GPU interconnects. The two best performing collective algorithms are \sysname-2 and \sysname-4. \sysname's collectives complete in nearly 80\% less time compared to both Ring and Tree algorithms with an ideal switch, despite incurring the MZI reconfiguration delay of \passage.

% Photonic connectivity with \passage challenges a key assumption of known algorithms like recursive doubling and halving: in photonic interconnects, each GPU can talk to multiple GPUs in a given round. A GPU on a \passage tile can simultaneously talk to as many other GPUs as the number of lasers on the tile. \sysname leverages this ability to generalize the recursive doubling/halving algorithms to \emph{recursive quadrupling/quartering}. The associated tradeoff is important. Splitting a GPU's total egress bandwidth across multiple wavelength-switched circuits allows 
% completing the collective in fewer rounds, reducing the $\alpha$ cost. However, using \emph{thinner} connections of less bandwidth instead of \emph{fatter} ones increases 
% the $\beta$ constant by $\frac{4}{3}$, leading to higher bandwidth cost. Since \passage has $16$ lasers, we can generalize this even further by splitting egress bandwidth from a GPU into 7 simultaneous transfers. However, not only this is complex, it also has diminishing returns. We call the doubling/halving and quadrupling/quartering algorithms as \sysname-2 and \sysname-4, respectively. Figure~\ref{fig:}

We evaluate the end-to-end benefit of training ML models across hundreds of GPUs connected in the \sysname architecture. We use the FlexFlow simulator to generate an optimal compute graph for a language model, BERT. We compare the training throughput of models trained on \sysname with the popular Ring algorithm over GPUs connected via an ideal switch. By comparing with ring on an ideal interconnect, we are comparing with the best possible performance \cite{topoopt_nsdi} can get. We use the state of the art 300 GB/s NVLink bandwidth per direction in our evaluation. We use the $\alpha$ value of 0.7 $\mu$s, derived by recent work for NVLinks~\cite{taccl}. For \sysname the $\alpha$ value is 0.7 + 3.7 $\mu$s to account for the reconfiguration delay. Figure~\ref{fig:perf}(a) shows that \sysname performs up to $1.7X$ better than the Ring algorithm.

BERT shows a high throughput improvement because the parallelization strategy has many \allreduce calls of small buffer sizes. Figure~\ref{fig:perf}(b) shows that small buffer \allreduce runtime is dominated by $\alpha$ cost especially at high bandwidths (300 GB/s), so \sysname algorithms 
perform better than the Ring algorithm which is only optimal in $\beta$ and linear in $\alpha$. \sysname's performance improves with the number of GPUs. \sysname is always better when the communication cost is dominated by $\alpha$ cost.

\section{Conclusion}
In conclusion, our results show significant advantages in ML training by integrating chip-to-chip photonic connectivity for multi-accelerator servers. Photonic connectivity with \sysname allows a GPU to communicate more effectively with multiple GPUs, unlike traditional collective algorithms for fixed network fabrics. Our results highlight the benefits of upgrading AI communications to use photonic interconnect solutions beyond the energy efficiency or the bandwidth alone.

\bibliographystyle{plain}
{\footnotesize
\bibliography{sample}
}
\end{document}